\documentclass[draft]{agujournal2018}
\usepackage{apacite}
\usepackage{bm}
\usepackage{url} 
\usepackage{pbox}
\usepackage{subfig}

\usepackage{ragged2e}

\draftfalse

 \journalname{arXiv.org}

\begin{document}
\justifying  
%
%


\title{Unification of visco-elastic wave equations}

%
%



\authors{Hao JIANG\affil{1}, Herv\'e CHAURIS\affil{1}}

\affiliation{1}{MINES ParisTech - PSL Research Univeristy}









\begin{keypoints}
\item Rheology
\item Fractional derivative model
\item Visco-elastic wave equations
\end{keypoints}

%
%


\begin{abstract}
Visco-elasticity is the essential ingredient for quantitative seismic imaging and geological interpretation in a number of contexts, such as in the presence of gas clouds. Decades of developments of numerical simulation of visco-elastic wave equations in seismology are mainly based on constant $Q$ model, leading to numerous different forms of time-domain visco-elastic wave equations. Based on rheological models, \cite{Emmerich1987} adopted the Generalized Maxwell body (GMB) to implement visco-elastic wave equations in time domain. \cite{Carcione1988a} incorporated the Generalized Zener body (GZB) into the time-domain visco-elastic wave equation. \cite{Moczo2005} proved that visco-elastic complex modulus based on GMB and GZB are equivalent. However, from the rheological point of view, this formalism can not incorporate the fractional visco-elastic wave equations based on the constant $Q$ model \citep{Kjartansson1979}. \cite{Mainardi2010} first mentioned that the constant $Q$ model is based on a fractional Scott-Blair model. The stress-strain relationship of the Scott-Blair model is between a spring and a dashpot. Therefore, we review the various visco-elastic wave equations in the text of seismology. Based on the stress-strain constitutive law of rheological models, we propose a unification way to describe the existed visco-elastic wave equations. The unification formalism indicates that each kind of visco-elastic wave equation is composed by the combination of basic rheological elements, e.g., GMB, GZB. 

In this paper, we gather knowledge usually available in separated papers in the fields of fractional calculus, rheology, mechanics, and seismology. By unification formalism, we can establish more clearly links between different approaches used in seismology.
\end{abstract}

\section{Introduction}
Linear visco-elasticity means that the stress is represented by a time relaxation function convoluted with strain. There are mainly 3 ways to describe the visco-elasticity. First, from the rheological point of view, the stress-strain relationship for a linear viscoelastic system can be represented by a spring-dashpot mechanical model (Figure~\ref{fig:rheology_md}a, b). Different combinations of springs and dashpots compose the family of visco-elasticity models. The general form of constitutive relation is expressed by an ordinary differential equation \citep{Fung1965}
\begin{equation}
\sum_{n=0}^{N} a_n \frac{d^n \sigma(t)}{dt^n} = \sum_{m=0}^{M} b_m \frac{d^m \epsilon(t)}{dt^m},
\label{eq:interger}
\end{equation}
where $n$ and $m$ are integer coefficients. Taking different orders of derivative and coefficients of $a_n$ and $b_m$ can obtain the constitutive relation between stress and strain, which represent different rheological models, such as: Maxwell body (MB), Kelvin body (KB), Zener body (ZB), Burgers body, Generalized Maxwell body (GMB)  or Generalized Zener Body (GZB) (Figure~\ref{fig:rheology_md}d, e, f, g, h, i). 
\begin{figure}[!htb]
  \centering 
  \fbox{
    \begin{minipage}{1.5 in} 
      \centering 
      \includegraphics[totalheight=2in]{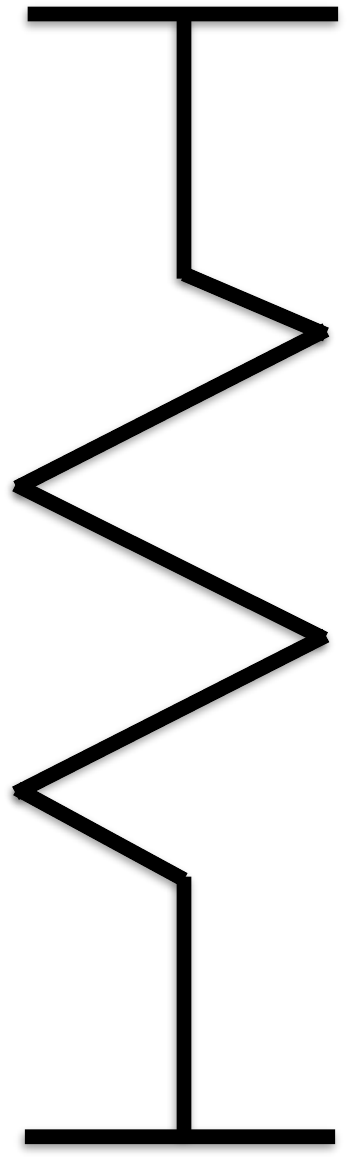} \\
      {(a): spring: $\sigma (t) = M_0  \epsilon(t)$} 
       \vspace{2.4pt}
      \label{fig:spring} 
      \end{minipage} }
    \fbox{
        \begin{minipage}{1.5 in} 
      \centering 
      \includegraphics[totalheight=2in]{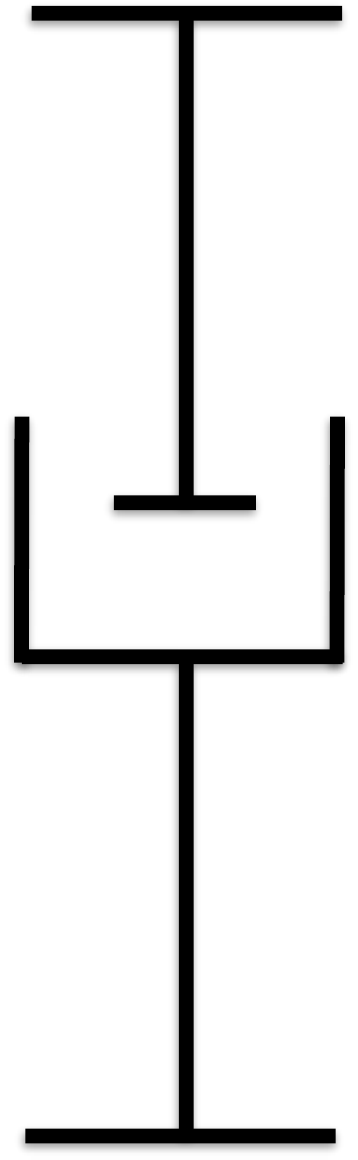} \\
      {(b): dashpot:  $\sigma (t) = \eta  \frac{\partial \epsilon (t)}{\partial t}$} 
      \label{fig:dashpot} 
    \end{minipage} } 
    \fbox{
        \begin{minipage}{1.5 in} 
      \centering 
      \includegraphics[totalheight=2in]{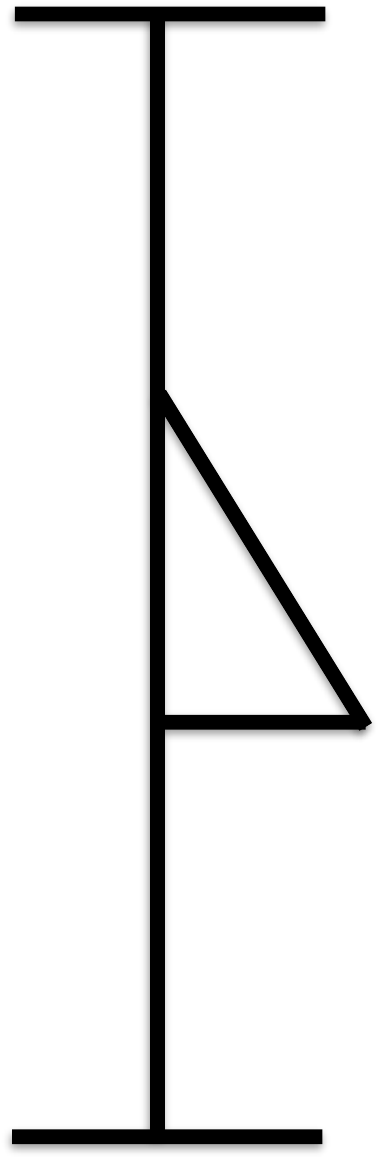} \\
      {(c): spring-pot: $\sigma (t) = G  \frac{\partial ^{\alpha}\epsilon (t)}{\partial ^{\alpha}t}$} 
      \label{fig:spring-pot} 
    \end{minipage} } 
    \fbox{
        \begin{minipage}{1.5 in} 
      \centering 
      \includegraphics[totalheight=2 in]{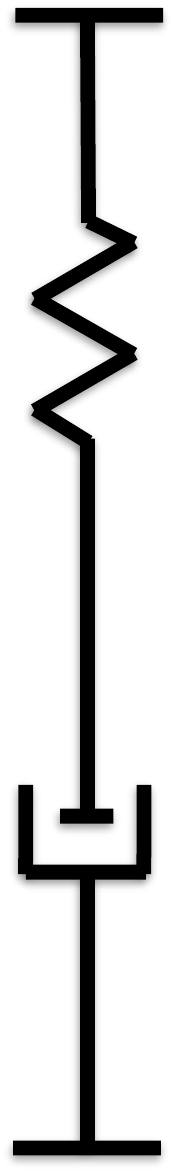} \\
      {(d): Maxwell: $\sigma (t) + \frac{\eta}{M_0} \frac{\partial \sigma (t)}{\partial t}= \eta  \frac{\partial \epsilon (t)}{\partial t}$}
      \vspace{37.5pt} 
      \label{fig:maxwell} 
    \end{minipage} } 
    \fbox{
        \begin{minipage}{1.5 in}
      \centering 
      \includegraphics[totalheight=2 in]{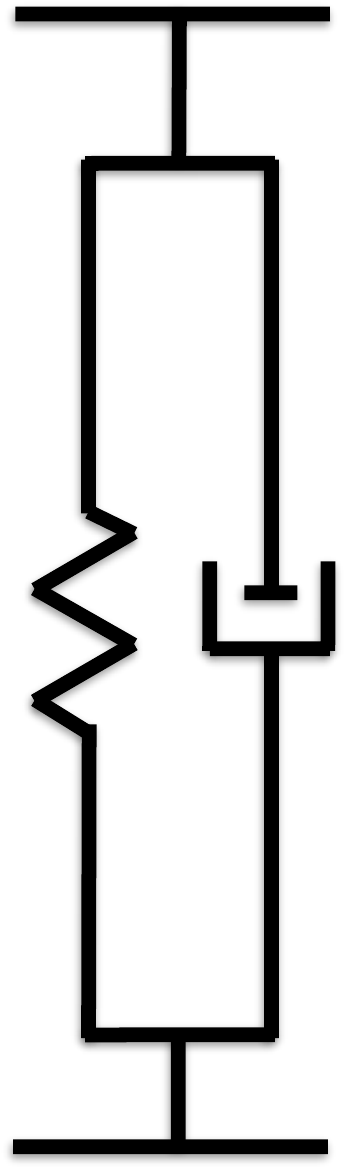} \\
      {(e): Kelvin: $\sigma (t) = M_0 \epsilon(t) + \eta  \frac{\partial \epsilon (t)}{\partial t}$} 
      \vspace{39.5pt}
      \label{fig:kelvin} 
    \end{minipage} } 
    \fbox{
        \begin{minipage}{1.5 in}
      \centering 
      \includegraphics[totalheight=2 in]{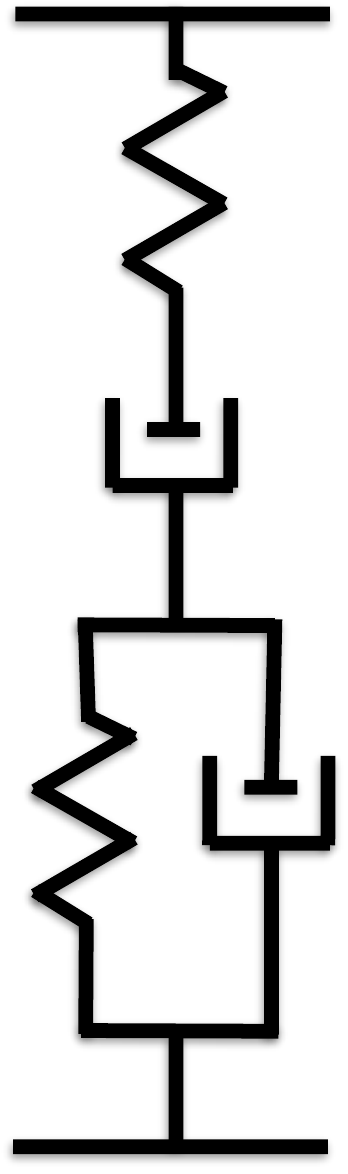} \\
      {(f): Burgers: $\sigma (t) + (\frac{\eta_1}{M_1}+\frac{\eta_1}{M_2}+\frac{\eta_2}{M_2}) \frac{\partial \sigma (t)}{\partial t} + \frac{\eta_1 \eta_2}{M_1 M_2}\frac{\partial^{2} \sigma (t)}{\partial^{2} t} = \eta_1  \frac{\partial \epsilon (t)}{\partial t} + \frac{\eta_1 \eta_2}{M_2}  \frac{\partial^{2} \epsilon (t)}{\partial^{2} t}$} 
       \vspace{7pt} 
    \label{fig:burgers} 
    \end{minipage} }
\end{figure}

\clearpage
\begin{figure}[!htb]
\ContinuedFloat     
    \fbox{
        \begin{minipage}{1.5 in}
      \centering 
      \includegraphics[totalheight=2in]{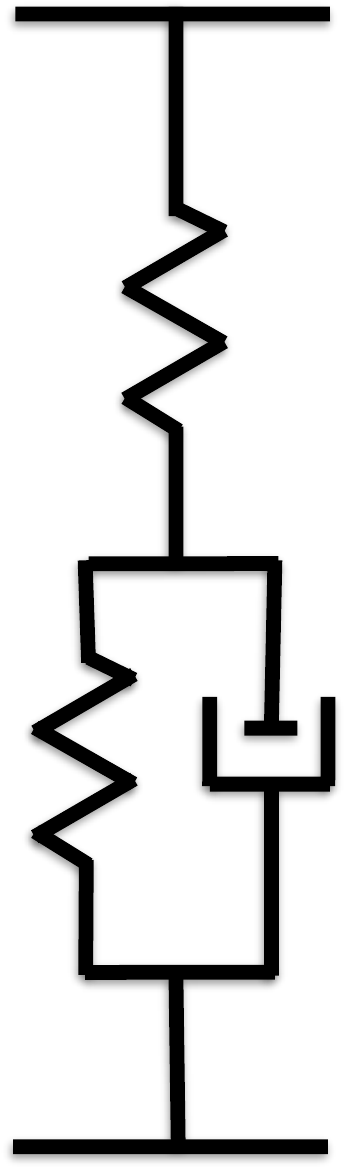} \\
      {(g): Zener: $\sigma (t) + \frac{\eta}{M_1+M_2}\frac{\partial \sigma (t)}{\partial t}= \frac{M_1M_2}{M_1+M_2}\epsilon(t) + \frac{M_1 \eta}{M_1+M_2}  \frac{\partial \epsilon (t)}{\partial t}$} 
      \label{fig:zener} 
    \end{minipage} } 
    \fbox{
        \begin{minipage}{1.5 in} 
      \centering 
      \includegraphics[totalheight=2in]{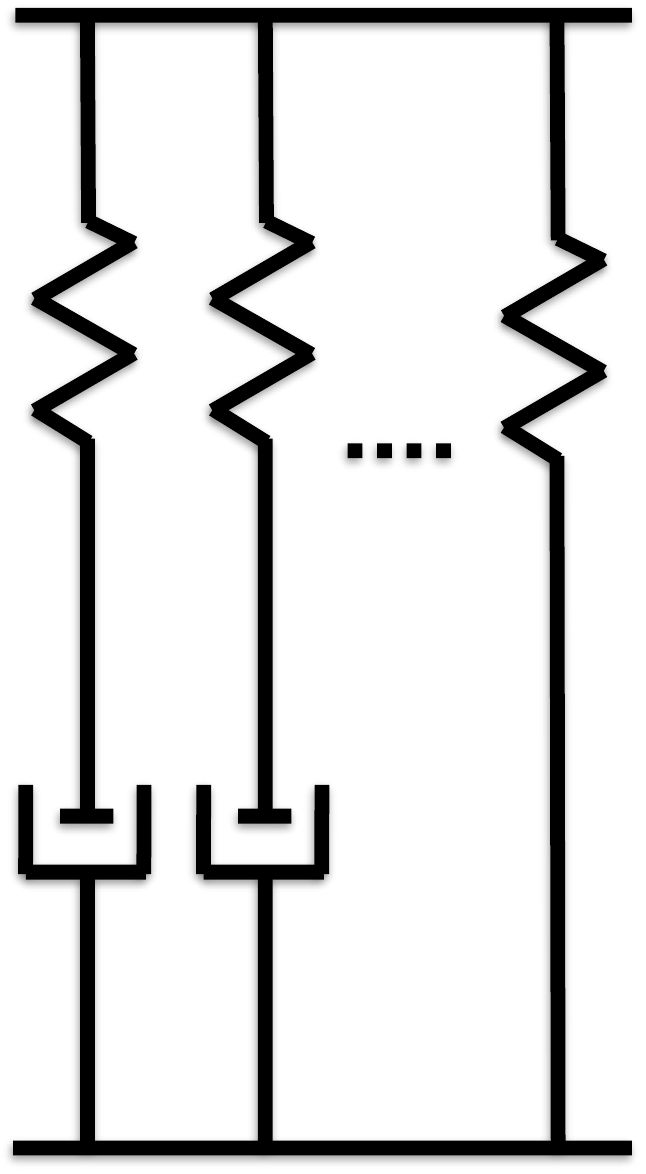} \\
      {(h): GMB: \\$\sigma (t) + \frac{\eta_l}{M_l}\frac{\partial \sigma (t)}{\partial t}= M_0 \epsilon(t) + (\frac{M_0 \eta_l}{M_l} + \eta_l)  \frac{\partial \epsilon (t)}{\partial t}$} 
      \label{fig:GMB} 
    \end{minipage} } 
    \fbox{
        \begin{minipage}{1.5 in} 
      \centering 
      \includegraphics[totalheight=2in]{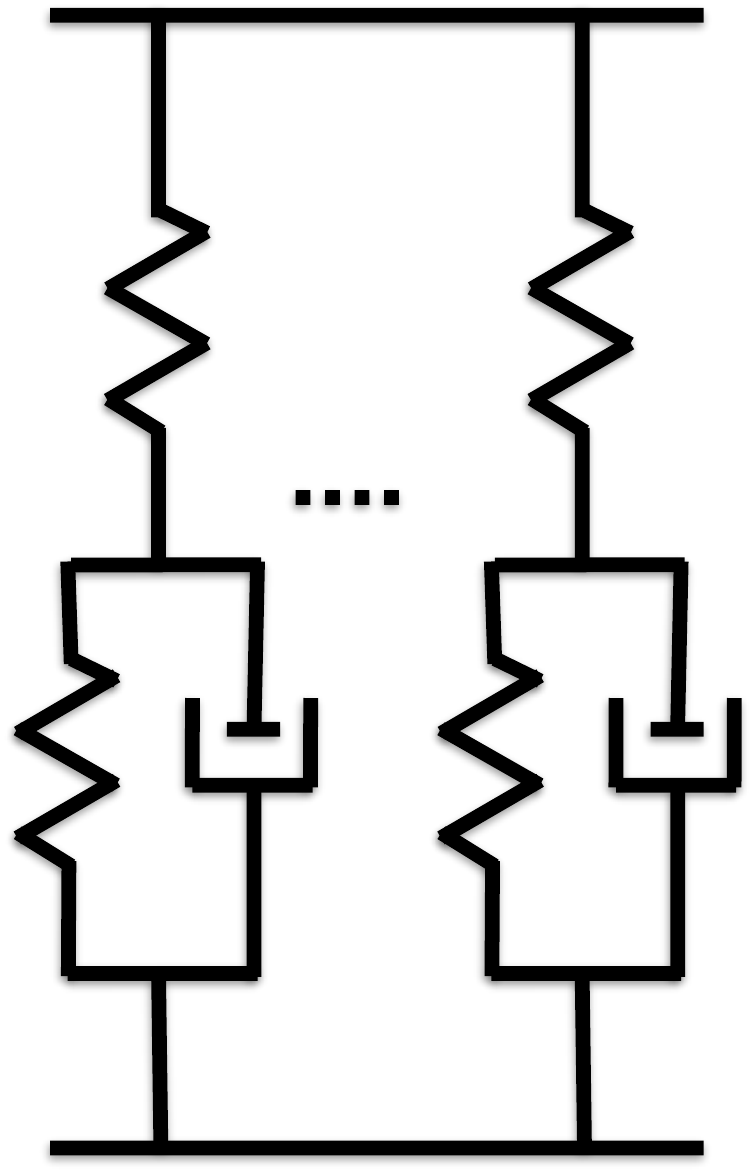} \\
      {(i): GZB: \\$\sigma(\omega) = \sum_{l=1}^{L} \frac{M_{1l}M_{2l} + i \omega \eta_l}{M_{1l}+M_{2l} + i \omega \eta_l}$} 
       \vspace{11pt}
      \label{fig:GZB} 
    \end{minipage} } 
    \caption{Elastic / rheological models and constitutive relations in the time domain, except GZB in the frequency domain due to the complexity in the time domain, $M_l$ is the elastic modulus of spring, $\eta _l$ is the viscosity coefficient and $G$ is the coefficient of spring-pot.} 
    \label{fig:rheology_md} 
\end{figure}

However, there is another kind of equation describing the relation between stress and strain. The idea behind is that the material properties are determined by various states between an elastic solid and a viscous fluid characterized as spring-pot (Figure:~\ref{fig:rheology_md}c), rather than a combination of an elastic and a viscous element represented by previous ordinary differential formula \citep{Gemant1938,Blair1947}. It can be represented by a fractional-order differential form \citep{Park2001}
\begin{equation}
\sum_{n=0}^{N} a_n \frac{d^{p_n} \sigma(t)}{dt^{p_n}} = \sum_{m=0}^{M} b_m \frac{d^{q_m} \epsilon(t)}{dt^{q_m}}, \hspace{10mm} 0 \leq p_n, q_m  \leq 1
\label{eq:fractional}
\end{equation}
where $p_n$ and $q_m$ are fractional derivative. In analogy with ordinary differential models, \cite{Mainardi2010} and \cite{Mainardi2011} displayed the fractional rheological models by replacing integer differential coefficients with fractional ones, e.g., fractional Maxwell model and fractional Zener body. Fractional differential models are widely applied in fields of materials and microstructure. 

Finally, another widely used theory to explain the visco-elasticity behaviors is called power law based on phenomenological models. The relaxation or creep time function has the shape of power function, which can well fit the observations of relaxation or creep behaviors of visco-elastic media \citep{Nutting1921}. The classic stress relaxation function gives as
\begin{equation}
\psi(t) =  M_{0} t^{-\alpha}, \hspace{10mm} 0 \leq \alpha \leq 1,
\label{eq:power_relaxation}
\end{equation}
where, $\psi(t)$ is the stress relaxation function, $M_0$ is the elastic modulus, and $\alpha$ is the power of the function. The stress is the time convolution between the relaxation function with the time derivative of strain in the time domain, leading to  
\begin{equation}
\sigma(t)=  M_0 \Gamma(1-\alpha) \frac{d^{\alpha} \epsilon(t)}{dt^{\alpha}},
\label{eq:power}
\end{equation}
where $\Gamma(1-\alpha)$ is the gamma function, $\alpha$ is the time differential coefficient. The first introduction in seismology of constant $Q$ model by \cite{Kjartansson1979} is based on the  phenomenological power law. The power law with fractional coefficient actually belongs to fractional derivative models \citep{Bagley1989,Park2001}, which is also applied for the visco-elastic simulation and inversion in seismology \citep{Hanyga2003,Ribodetti2004}.  

After analyzing the three formulas, we can write Equations~\ref{eq:interger}, \ref{eq:fractional} and \ref{eq:power} as a general form 
\begin{equation}
\sum_{n=0}^{N} a^l_n \frac{d^{p_n} \sigma(t)}{dt^{p_n}} = \sum_{m=0}^{M} b^l_m \frac{d^{q_m} \epsilon(t)}{dt^{q_m}}, \hspace{10mm}1 \leq l \leq L
\label{eq:general}
\end{equation}
where $l$ is associated to a rheological model among total number $L$, $p_n = n$ and $q_m = m$ correspond to ordinary order representing, e.g., GZB, $0 \leq p_n, q_m  \leq 1$ corresponds to fractional order representing fractional derivative models, and models with $N=0, M=0$ can be explained by the classic power law.

To clarify the relations of different theories and the connections between different visco-elastic wave equations, we review different rheological models and applications in seismology. Most commonly used vicso-elastic wave equations obtained by the combinations of springs and dashpots, e.g. GZB and GMB models, are based on nearly constant $Q$ model. \cite{Kjartansson1979} constant $Q$ model and associated visco-elastic wave equation are based on power law. Other applications in seismology are based on power law \citep{Hanyga2003,Ribodetti2004,Wang2016}. The power law belongs to the family of fractional derivative models.

Two prominent papers should be mentioned. The first one is written by \cite{Carcione2011}, who gave different mechanical visco-elastic models and relevant wave equations. Those mechanical models can be obtained from ordinary differential equation (Equation~\ref{eq:interger}). The second is written by \cite{Mainardi2010}, who mentioned that \cite{Kjartansson1979} constant $Q$ is a specific example of fractional differential models called Scott-Blair model. Those different visco-elastic wave equations and their rheological models need a unification formalism to be represented in exploration seismology. Thus we combine the three equations (Equation~\ref{eq:interger}, \ref{eq:fractional} \ref{eq:power}) into a unification formula (Equation~\ref{eq:general}). This unification formula is a easy way to identify each specific form of visco-elastic wave equations and the related rheological models (Table~\ref{tbl:review}). Furthermore, it is a way to obtain other visco-elastic wave equation to describe more complicated attenuation behaviors, which is not limited to seismology. We begin from the unified stress-strain relationship to obtain the general visco-elastic wave equation, and give specific examples with their associated rheological models and visco-ealstic wave equations.
\begin{table}[]
\centering
 \caption{A review of visco-elasticity, $\Downarrow$ \&$ \Leftarrow$ lead to the contents in the arrow direction.}
\begin{tabular}{c|c|c|c}
\hline
In seismology                                                         & \multicolumn{3}{c}{Viscoelasticiy}                                                                                                                                                                                                                                                        \\ \hline
\begin{tabular}[c]{@{}c@{}}Existing\\ equations\end{tabular}           & \begin{tabular}[c]{@{}c@{}}ZB, MB, GZB, GMB ...\\ (Nearly constant $Q$)\\ \cite{Emmerich1987}\\ \cite{Carcione1988a}\\ $\Downarrow$ \end{tabular}               & \begin{tabular}[c]{@{}c@{}}Constant $Q$\\ \cite{Kjartansson1979} \\$\Downarrow$ \end{tabular}                 & \begin{tabular}[c]{@{}c@{}}$\Leftarrow$ \hspace{3mm}Power law\\ \cite{Bagley1989} \\ $\Downarrow$ \end{tabular}                           
   \\ \hline
\begin{tabular}[c]{@{}c@{}}Differential \\ equation\end{tabular}          & \begin{tabular}[c]{@{}c@{}} Ordinary order\\ \cite{Moczo2005} \\ \cite{Carcione2011}\\$\Downarrow$\end{tabular}                                     & \multicolumn{2}{c}{\begin{tabular}[c]{@{}c@{}} Fractional order\\ \cite{Mainardi2010}\\ $\Downarrow$\end{tabular}}                                                                         \\ \hline
\begin{tabular}[c]{@{}c@{}}Stress-strain \\ relationship\end{tabular} & \multicolumn{3}{c}{\begin{tabular}[c]{@{}c@{}}Unification formalism (\textbf{Here})\\ $\Downarrow$\end{tabular}}                                                                                                                                                                                                                                          \\ \hline\hline
\begin{tabular}[c]{@{}c@{}}Specific\\  example\end{tabular}           & \begin{tabular}[c]{@{}c@{}}One spring with \\ $L$ pairs of\\ M~=~N~=~1,\\ $p_0$~=~0, $q_0$~=~0 \\ and $p_1$~=~1, $q_1$~=~1\end{tabular} & \begin{tabular}[c]{@{}c@{}}M~=~N~=~0,\\ $p_0$~=~0, $q_0$~=~$\alpha$ \\ and $L$~=~1\end{tabular} & \begin{tabular}[c]{@{}c@{}}Others, e.g., \\ fractional Zener \\M~=~N~=~1,\\ $p_0$~=~0, $q_0$~=~0 \\ and $p_1$~=~$\alpha$, $q_1$~=~$\beta$ \\ $L$~=~1 \cite{Mainardi2010}\end{tabular}                    \\ \hline
$Q$ model                                                               & Nearly constant $Q$                                                                                               & Constant $Q$                                                                & Others                                                                                       \\ \hline
\begin{tabular}[c]{@{}c@{}}Rheological \\ elements\end{tabular}       & Spring, dashpot                                                                                                 &  \begin{tabular}[c]{@{}c@{}}Spring-pot  \\ (Scott-Blair model)\end{tabular}                                                             & \begin{tabular}[c]{@{}c@{}}Combinations of \\ Spring, dashpot \\ and spring-pot\end{tabular} \\ \hline
\end{tabular}
 \label{tbl:review}
\end{table}
\section{Complex modulus and visco-elastic wave equations}
For $L$ pairs of rheological models in parallel, the stress and strain relation in frequency domain is given by
\begin{equation}
\sum_{n=0}^{N} a^l_n (i \omega)^{p_n} \sigma (\omega) = \sum_{m=0}^{M} b^l_m \ (i \omega)^{q_m} \epsilon (\omega), \hspace{10mm}1 \leq l \leq L.
\end{equation}
Then the complex modulus \citep{Bland1960,White1965} reads 
\begin{equation}
M(\omega) = \sum_{l=1}^{L} \frac{\sum_{m=0}^{M} b^l_m \ (i \omega)^{q_m} }{\sum_{n=0}^{N} a^l_n (i \omega)^{p_n}}. 
\label{eq:complex_M}
\end{equation}
The imaginary part corresponds to energy loss. The $Q$ factor \citep{Boit1954,Knopoff1956,Fung1965,O'Connell1978,Ben1981} is defined as
\begin{equation}
\frac{1}{Q} = \frac{\Im[M(\omega)]}{\Re[M(\omega)]}.
\label{eq:Q_define}
\end{equation}
From the constant $Q$ or nearly constant $Q$ model based on the observations of seismic waves and rock experiments \citep{McDonal1958}, one first needs to have the proper number for $M$, $N$, $p_n$ and $q_m$ to dertermine the basic rheological element. Then the objective is to determine the $a_n$ and $b_m$ corresponding to elastic modulus and viscosity coefficients and $L$ to obtain constant $Q$ within the frequency range \citep{Liu1976}. 
With the constant $Q$ optimization, then the general form of visco-ealstic wave equation is obtained with the complex modulus in the frequency domain. The general form of visco-acoustic wave equation with constant density is
\begin{equation}
(i \omega)^2 p - \frac{1}{\rho} M(\omega) \nabla ^{2} p =0,
\end{equation}
where $p$ is the pressure and $\rho$ is the density. The complex modulus $M(\omega)$ leads to visco-elastic effects including dispersion and dissipation. Different forms of $M(\omega)$ are obtained from different ways, e.g., phenomenological observation of relaxation or creep function, leading to the existed various forms of visco-elastic wave equations. From a rheological point of view, $M(\omega)$ can be derived from various rheological models. Thus we will derive visco-acoustic wave equations from the unification formalism including nearly constant $Q$ based on GMB and \cite{Kjartansson1979} constant $Q$. Compared with the visco-acoustic wave equation, visco-elastic wave equation only requires to introduce $Q_p$ for P wave and $Q_s$ for S wave separately. Both $Q_p$ and $Q_s$ are determined by the same way. Thus we only illustrate the visco-acoustic wave equations from the unification formalism based on different rheological models, instead of visco-elastic wave equations. 
 
\section{Example I: Nearly constant $Q$ and associated wave equations}
Nearly constant $Q$ model is approximated by several rheological models like GZB and GMB. \cite{Emmerich1987} first introduced the GMB rheological model to obtain nearly constant $Q$. \cite{Carcione1988a, Carcione1988b} incorporated the GZB into the time-domain visco-elastic wave equation. \cite{Moczo2005} proved the equivalence between GZB and GMB and put forward a material-independent formula. However, this material-independent formula is still based on the ordinary differential constitutive relation (Equation~\ref{eq:interger}). We choose the integer values to obtain nearly constant $Q$ from equation (Equation~\ref{eq:general}) and visco-acoustic wave equations. 

The GMB is composed by a spring and $L$ pairs of Maxwell body in parallel.
A spring is represented as $M=N=0$, $p_0 = 0$, $q_0 = 0$: the constitutive relation is
\begin{equation}
\sigma (\omega) = M_0  \epsilon(\omega). 
\end{equation}
A Maxwell body is represented by $M=N=1$, $p_0 = 0$, $q_0 = 0$ and $p_1 = 1$, $q_1 = 1$: the constitutive relation leads to
\begin{equation}
\sigma (t) + \frac{\eta_l}{M_l} \sigma^{(1)}(t)= M_l \frac{\eta_l}{M_l}  \epsilon^{(1)}(t). 
\end{equation}
In frequency domain is expressed as
\begin{equation}
\sigma (\omega) + \frac{i \omega}{\omega_l} \sigma(\omega)= M_l  \frac{i \omega}{\omega_l}  \epsilon(\omega). 
\end{equation}
$\omega_l$ is equal to $1 / t_l$, in which $t_l$ is referred to the relaxation time from the stress relaxation function of Maxwell body $\psi(t) = M_l$exp$(-t/t_l)H(t)$. $t_l$ is defined as $\eta _l / M_l$. 
Complex modulus for $l$ Maxwell body is 
\begin{equation}
M^{l} (\omega)= \frac{M_l i \omega}{\omega_l + i \omega}. 
\end{equation}
Therefore the total complex modulus of GMB is 
\begin{eqnarray}
M (\omega) &= & M_0 + \sum_{l=1}^{L} \frac{M_l i \omega}{\omega_l + i \omega}, \nonumber \\
                   &=&  M_0 + \delta M \sum_{l=1}^{L} \frac{\alpha_l i \omega}{\omega_l + i \omega}, 
\end{eqnarray}
with  $\alpha_l \delta M = M_l$. $M_l$ is the elastic modulus of each Maxwell body and $M_0$ is the elastic modulus of the spring. Based on rheological models, one needs to optimize $a_n$ and $b_m$ corresponding to elastic modulus and viscosity coefficients and $L$ to obtain constant $Q$ within the frequency range (Equation~\ref{eq:complex_M}, \ref{eq:Q_define}). Here $a_n$ and $b_m$ have been parameterized as $\omega_l$ and $\alpha_l$. $\omega_l$ is usually determined logarithmically over the frequency range \citep{Blanch1995}. Different $\alpha_l$ optimizes different $Q$ values. 

The wave-equation reads
\begin{equation}
(i \omega)^2 p - \frac{1}{\rho} (M_0 + \delta M \sum_{l=1}^{L} \frac{\alpha_l i \omega}{\omega_l + i \omega}) \nabla ^{2} p =0
\end{equation}
Therefore the form in time domain is 
\begin{eqnarray}
\frac{\partial{p}}{\partial{t}} + \frac{1}{\rho} (M_0+ \delta M \sum_{l=1}^{L} \alpha_l) (\nabla \cdot \bm{v}) + \sum_{l=1}^{L}r^{l}_{p} &=& f(x_{s},t), \nonumber \\ 
\frac{\partial{\bm{v}}}{\partial t} +  \frac{1}{\rho} \nabla p &=& 0,\nonumber \\
\frac{\partial{r^{l}_{p}}}{\partial t} + \omega_l(r^{l}_{p}+ \delta M \alpha_l \nabla \cdot \bm{v})&=&0, \hspace{10mm}l=\in[1, L]. 
\end{eqnarray}
$\bm{v} $ is the particle velocity vector. $r^{l}_{p}$ is the memory variables. 

From the general formula of stress-strain relationship, we get the visco-acoustic wave-equation based on GMB rheological model. The visco-equation can be obtained by GZB rheological model. Next section, we introduce the \cite{Kjartansson1979} constant $Q$ visco-acoustic wave equation the under frame of the unification formalism.



\section{Example II: Kjartansson constant $Q$ and associated wave equations}
\cite{Kjartansson1979} gives a relaxation function and complex modulus defined as 
\begin{equation}
M(\omega) = M_0 (\frac{i \omega}{\omega_0})^{2 \gamma}, \hspace{5mm}  0 < 2 \gamma < 1
\label{eq:Q_kjar_chapter3}
\end{equation}
where $\omega_0$ is the reference frequency. The complex modulus is derived from a power creep function. \cite{Mainardi2010} pointed out this form can be obtained from the rheological model called Scott-Blair model. In fact, the Scott-Blair model called by \cite{Mainardi2010} is a spring-pot model (Figure~\ref{fig:rheology_md}). The stress of a spring-pot is linear, with the fractional derivative of strain. Taking $M=N=0$, $p_0 = 0$, $q_0 = \alpha $ and $L=1$ in Equation~\ref{eq:general}, the constitutive relation turns into
\begin{equation}
\sigma (t) = M_0 (\frac{\eta_0}{M_0})^{\alpha} \epsilon^{(\alpha)}(t)
\end{equation}
by choosing the $b_0 / a_0$ as $M_0(\eta_0/M_0)^{\alpha}$. $M_0$ stands for the elastic modulus, $\eta$ is the viscous coefficient of a dashpot. Defining $t_0= 1/ \omega_0$ as $\eta _0 / M_0$, $t_0$ is referred as the relaxation time. The stress-strain relation in the frequency domain is
\begin{equation}
\sigma (\omega) = M_0 (\frac{1}{\omega_0})^{\alpha} (i \omega)^{\alpha} \epsilon(\omega). 
\end{equation}
Thus the complex modulus is defined as
\begin{equation}
M(\omega) = M_0 (\frac{i \omega}{\omega_0})^{\alpha} =  M_0 (\frac{i \omega}{\omega_0})^{2 \gamma},
\end{equation}
with $\alpha= 2 \gamma$, we have the complex modulus equivalent to Equation~\ref{eq:Q_kjar_chapter3} which is obtained through the power law. The parameter $\alpha$ depends the $Q$ value.
The wave equation is derived by introducing the complex modulus as
\begin{equation}
(i \omega)^2 p - \frac{1}{\rho} M_0 (\frac{i \omega}{\omega_0})^{2 \gamma} \nabla ^{2} p =0.
\end{equation}
Combing the $(i \omega)^{2 \gamma}$ with $(i \omega)^2$ and transferring into time domain, the visco-acoustic wave equation reads  
\begin{equation}
\frac{\partial^{2-2 \gamma} p}{\partial t^{2- 2 \gamma}} - \frac{M_0}{\rho} \omega_0^{-2 \gamma} \nabla ^{2} p = 0. 
\end{equation}
This function is consistent with the result of \cite{Carcione2002}. This form of visco-acoustic wave equation is obtained with a spring-pot rheological model. The constant $Q$ visco-acoustic wave equation in time domain involves with time-domain fractional derivatives. It is hard to be solved by finite difference method. Thus the commonly applied visco-acoustic wave equation is based on GMB / GZB rheological models by introducing memory variables instead of solving fractional time derivative \citep{Emmerich1987,Carcione1988a}. 

\section{Conclusions}
The general constitutive relation of stress and strain combines ordinary differential system, fractional order differential system and power law together. From the constitutive relation, the unification formalism is a way to describe and summarize visco-elastic wave equations in seismology. Depending on the different strategies of optimizing constant $Q$, the rheological models of visco-elastic wave equations contain different combinations of basic rheological elements with springs, dashpots and spring-pots. The widely applied visco-elastic wave equations in seismology can be characterized as different specific cases. Furthermore, it is a way to obtain other kinds of visco-elastic wave equations, e.g. based on fractional Zener model. Those visco-elastic wave equations can be summarized under the unification formalism. 

 \bibliography{haojiang.bib}

%




\end{document}